\documentclass[12pt,twoside]{article}
\usepackage{amsmath}
\numberwithin{equation}{section}
\newcommand{\rd}[1]{\mathop{\mathrm{d}#1}}

\newcommand{\fract}[2]{{\textstyle\frac{#1}{#2}}}

\newcommand{\CS}{Chern-Simons}
\newcommand{\CSt}{Chern-Simons term}

\newcommand{\nc}{noncommutative}

\newcommand{\cc}{consistency condition}
\newcommand{\swm}{Seiberg-Witten map}
\newcommand{\tn}{transformation}
\newcommand{\oc}{1-cocycle}
\newcommand{\hL}{\ensuremath\hat\Lambda}
\newcommand{\hA}{\ensuremath\hat A}

\newcommand{\numeq}[2]{\begin{equation}
#2
\label{#1}
\end{equation}}
\newcommand{\refeq}[1]{(\ref{#1})}

\newcommand{\ts}{\thinspace}
\newcommand{\citer}[1]{\ts\cite{#1}}

\let\epsilon\varepsilon
\let\phi\varphi

\begin{document}
 
\title{Noncommutative 1-cocycle in the Seiberg-Witten map}
\author{R. Jackiw\\
\small\it Center for Theoretical Physics\\ 
\small\it Massachusetts Institute of Technology\\ 
\small\it Cambridge, MA 02139-4307\\[2ex]
S.-Y. Pi\\
\small\it Physics Department\\ 
\small\it Boston University \\
\small\it Boston MA 02215, USA}

\date{January 2002}
%\\
\date{\small BUHEP-02-05\quad
MIT-CTP-3241\quad hep-th/0201251\\
\footnotesize Typeset in \LaTeX\ by M. Stock}
\maketitle

\abstract{\noindent We show that the \swm\ for a \nc\ gauge theory involves a \nc\ \oc.
The cocycle condition enforces a consistency requirement, which has been
previously derived.}

\thispagestyle{empty}

\section{Introduction}\label{s1}
The chiral anomaly\citer{r1} and its \cc\citer{r2} have been
given a cohomological formulation in terms of infinitesimal gauge
\tn s by Stora and Zumino\citer{r3}. Alternatively, Faddeev, Shatashvili, and
Mickelsson\citer{r4} used finite \tn s to construct on the gauge group cocycles,
which in infinitesimal form reproduce the anomaly, anomalous
commutators\citer{r5} and the \cc.

Apparently a similar story can be told for the \swm\ in a \nc\ gauge
theory\citer{r6}. A \cc\ has been identified by Jur\v co et~al.\citer{r7},  and a
cohomological approach, in terms of infinitesimal quantities, has been
constructed by Brace et~al.\citer{r8}.

Here we continue this parallelism with anomaly theory by considering the \swm\
in terms of finite \tn s, and thereby construct a \oc, which is noncommuting. 

In Section~\ref{s2} we recall the definition and properties of a \oc, which is then
extracted from the \swm\ in Section~\ref{s3}.

\section{\oc\ (review)}\label{s2}
Consider a group of \tn s $\{g\}$, ($g_1 g_2 = g_{12}$) on some
coordinates~$\xi$: $\xi\to\xi^g$. Let these \tn s be implemented on functions of
$\xi$, $\Psi(\xi)$, by some operation $U(g)$. The simplest action of $U$ on
$\Psi$ would be $U(g)\Psi(\xi) = \Psi(\xi^g)$. But this can be  generalized by
allowing a factor to appear: 
\numeq{e2.1}{
\begin{aligned} 
U(g)\Psi(\xi) &= C(\xi,g) \Psi(\xi^g)\\
C(\xi,I) &= 1\ .
\end{aligned}
}
If the $U$'s obey the group composition law
\numeq{e2.2}{
U(g_1)U(g_2) = U(g_{12})
}
then $C$ must satisfy a condition,
which follows by effecting a second transformation on~\refeq{e2.1}:
\numeq{e2.3}{
C(\xi,g_{12}) = C(\xi,g_1) C(\xi^{g_1}, g_2)\ .
}

A quantity that depends on one group element  $(g)$ and possibly on  the
coordinates $(\xi)$ is called a 1-cochain. If it also satisfies \refeq{e2.3} it is a \oc.
When a \oc\ can be written as
\numeq{e2.4}{
C(\xi,g) = C_0^{-1}(\xi) C_0(\xi^g)
}
it is trivial: \refeq{e2.4} certainly satisfies \refeq{e2.3}, but $C$ can be removed
from \refeq{e2.1} by replacing $\Psi$ by $C_0\Psi$ and $U$ by $C_0 U
C_0^{-1}$. A trivial cocycle is called a coboundary. When a 1-cochain is written
in exponential form
\numeq{e2.5}{
\begin{gathered} 
C(\xi,g) = \exp\bigl(-i \gamma(\xi,g)\bigr)\\
\gamma(\xi,I) = 0 \mod
2\pi (\mathrm{integer})
\end{gathered}  
}
the \oc\ condition \refeq{e2.3} may be presented as 
\numeq{e2.6}{
\gamma(\xi^{g_1}, g_2) - \gamma(\xi, g_{12}) + \gamma(\xi, g_1) = 0 \mod
2\pi (\mathrm{integer}) 
}
and the \oc\ is a trivial coboundary when
\numeq{e2.7}{
\gamma(\xi, g) = \gamma_0(\xi^g) - \gamma_0(\xi)\ .
}

[Generalizations of the above include 2-cocycles (when the composition law for
$U$ acquires a modification) and 3-cocycles (when the composition law for
$U$ is nonassociative)\citer{r9}.]

In the application  to anomalies, $\xi$ is the vector potential and $g$ is a  gauge \tn.
Then the anomaly is the infinitesimal portion of~$\gamma$ and the \cc\ is the
infinitesimal version of the  \oc\ condition 
\refeq{e2.6}. 

\section{\oc\ (\swm)}\label{s3}
The \swm\ arises from the requirement that a noncommutative gauge potential
$\hA$, viewed as a function of the commutative gauge potential~$A$, be
stable against gauge \tn s, in the sense that\citer{r6}
\numeq{e3.1}{
\hat A(A) + D\hL(A,\alpha) = \hat A(A+D\alpha)\ .
}
Here $\hL$ and  $\alpha$ are infinitesimal parameters of a
noncommutative and commutative gauge \tn, respectively:
\numeq{e3.2}{
\begin{aligned}
 D\hL &= \rd{\hL} - i[\hat A, \hL]_\star\\
D\alpha &= \rd\alpha - i[ A, \alpha]\ .
\end{aligned}
}
As usual, the star product, involving the noncommutativity parameter~$\theta^{ij}$,
forms the star commutator $[\hat A, \hL]_\star = 
\hat A\star\hL -  \hL \star\hat A$. $\hL$ depends on $A$  and $\alpha$
with
\numeq{e3.3}{
\hL(A,0) = 0\ .
}
($\hA$ and $\hL$ also depend on $\theta$, but this will not be indicated
explicitly.)

When the \swm\ is extended to additional fields, transforming with the
fundamental representation of the gauge group, a \cc\ has been derived. Define
\numeq{e3.4}{
\hL(A,\alpha) = \Lambda_\alpha(A) + \Lambda_\alpha^{(2)}(A) + \cdots\ .
}
$\Lambda_\alpha(A)$ is the portion of $\hL(A,\alpha)$ that is linear
in~$\alpha$; $\Lambda_\alpha^{(2)}(A)$ is the quadratic part; in view of
\refeq{e3.3} there is no $\alpha$-independent contribution. The \cc\ then
reads\citer{r7}
\numeq{e3.5}{
\delta_\alpha \Lambda_\beta - \delta_\beta \Lambda_\alpha - 
i[\Lambda_\alpha,\Lambda_\beta]_\star + i \Lambda_{[\alpha,\beta]} = 0\ .
}
Here
\numeq{e3.6}{
\delta_\alpha \Lambda_\beta = \Lambda_\beta(A+D\alpha) -
\Lambda_\beta \ .
}
[When $\Lambda_\alpha$, $\Lambda_\alpha^{(2)}$ are written without an
argument, the missing argument is understood to be~$A$; other arguments are
indicated explicitly as in \refeq{e3.6}.]

Rather than considering the response to infinitesimal \tn s as in \refeq{e3.1}, we
use finite gauge \tn s and posit the finite version of~\refeq{e3.1}:
\numeq{e3.7}{
\hA^G(A) = \hA(A^g)\ .
}
Here $A^g$ is the commutative gauge \tn\ of the commutative
potential~$A$:
\numeq{e3.8}{
A^g = g^{-1} A g + g^{-1}i \rd g\ .
}
Similarly $\hA^G$ is the noncommutative gauge \tn\ of the noncommutative
potential~$\hA$:
\numeq{e3.9}{
\hA^G = G^{-1}\star A\star  G + G^{-1}\star i \rd G\ .
}
$G$ depends on $A$ and $g$. 

Consider now $\hA(A^{g_1g_2})$. We may view $A^{g_1}$ as a new gauge
potential $A'$ and $A^{g_1g_2}$ as the $g_2$-transformation of~$A'$. Then
\refeq{e3.7} implies
\begin{subequations}\label{e3.10}
\begin{align}
\hA(A^{g_1g_2}) 
&= \hA(A^{\prime g_2}) 
= G^{-1}(A',g_2)\star \bigl(\hA(A')\star + i \rd{\null}\bigr)
G(A',g_2)\label{e3.10a}\\ 
&= G^{-1}(A^{g_1}, g_2)\star \Bigl(
G^{-1}(A,g_1)\star \bigl(\hA(A)\star + i \rd{\null}\bigr) G(A,g_1)\star +
i \rd{\null} \Bigr) G(A^{g_1}, g_2)\notag\\
&= \bigl( G(A,g_1)\star G(A^{g_1}, g_2)\bigr)^{-1} \star 
\bigl(\hA(A)\star + i \rd{\null}\bigr) \bigl( G(A,g_1)\star G(A^{g_1}, g_2)\bigr)\
.\notag\\
\intertext{Alternatively $A^{g_1g_2}$ is also the $g_{12}$ transform of~$A$.
Then \protect\refeq{e3.7} gives}
\hA(A^{g_1g_2}) & = \hA(A^{g_{12}}) 
= G^{-1}(A,g_{12})\star \bigl(\hA(A)\star + i \rd{\null}\bigr)
G(A,g_{12})\ . \label{e3.10b}
\end{align}
\end{subequations}
Comparing the two results in the equation
\numeq{e3.11}{
G(A,g_{12}) = G(A,g_1)\star G(A^{g_1}, g_2)\ .
}
This is the same as the \oc\ condition \refeq{e2.4}, except that star
multiplication has replaced ordinary multiplication, namely, the \oc\ $G$ is
noncommutative. 

The \oc\ would be trivial if it were given, analogously to \refeq{e2.4}, by 
\numeq{e3.12}{
G(A,g) = G_0^{-1}(A)\star G_0(A^g)
}
which certainly satisfies \refeq{e3.11}. Moreover, using this trivial cocycle in
\refeq{e3.7}--\refeq{e3.9} implies
\numeq{e3.13}{
G_0(A)\star \bigl(\hA(A)\star+i \rd{\null}\bigr) G_0^{-1}(A) 
= G_0(A^g)\star \bigl(\hA(A^g)\star+i \rd{\null}\bigr) G_0^{-1}(A^g)\ .
}
This states that the transform of $\hA(A)$ with the noncommuting gauge \tn\
$G_0^{-1}(A)$ results in a quantity that is invariant against commuting gauge
\tn s of~$A$. Presumably this can only be true if $\hA$ is a pure gauge: $\hA=
G_0^{-1}\star i \rd{G_0}$, or a gauge \tn\ (by~$G_0$) of an $A$-independent
noncommuting potential $\hA_0$:
\numeq{e3.14}{
\hA(A) = G_0^{-1}(A) \star \hA_0 \star G_0(A)  
+ G_0^{-1}(A) \star  i \rd{G_0(A)}\ . 
}

$G_0$ also parameterizes an ambiguity in solutions to \refeq{e3.11}: If $G(A,g)$
solves \refeq{e3.11}, so does $G_0^{-1}(A) \star G(A,g)\star G_0(A^g)$.
When this form for the cocycle/gauge \tn\ is used in \refeq{e3.7} and terms are
rearranged, we are left with
\numeq{e3.15n}{
\begin{gathered}
G_0  (A^g)\star \hA(A^g)\star G_0^{-1}(A^g) -  i \rd{G_0} (A^g)\star G_0^{-1} (A^g)\\
= G^{-1} (A,g) \star \Bigl(
G_0  (A)\star \hA(A)\star G_0^{-1}(A) -  i \rd{G_0} (A)\star G_0^{-1} (A) \star G(A,g)
\Bigr)\\
{}+ G^{-1} (A,g)\star  i \rd{G(A,g)}\ .
\end{gathered}
}
This equation demonstrates 
%that if $G(A,g)$ is the appropriate cocycle/gauge \tn\  for 
%$\hA(A)$, in the sense of \refeq{e3.7}--\refeq{e3.9}, then the same cocycle does the job
%for an arbitrary gauge transform of $\hA(A)$ [by $G_0^{-1}(A)$].
the gauge covariance of the formalism.

In analogy to \refeq{e2.5} and consistent with \refeq{e3.1}, \refeq{e3.7},
\refeq{e3.9}, an exponential form for $G$ may still be used: 
\numeq{e3.15}{
G= e^{-i\hL} \ .
}
But a formula analogous to \refeq{e2.6} cannot be established because for
noncommuting quantities products of exponentials are not simply exponentials
of summed exponents. Nevertheless, the expression \refeq{e3.15} may be used to
derive the consistency condition \refeq{e3.5} from \refeq{e3.11}. We set 
$g_1=e^{-i\alpha}$, $g_2=e^{-i\beta}$, and label~$\hL$ by generator~$\alpha$
or~$\beta$. Thus
\begin{subequations}\label{e3.16}
\numeq{e3.16a}{
G(A, g_1) = I - i\hL(A,\alpha) - \fract12 \hL(A,\alpha)\star \hL(A,\alpha) + 
\cdots\ .
 }
It will be necessary to work to quadratic order, so according to \refeq{e3.4} we
have
\numeq{e3.16b}{
G(A, g_1) = I - i\Lambda_\alpha  - \fract12 \Lambda_\alpha\star
\Lambda_\alpha - i\Lambda_\alpha^{(2)}\  .
 }
Consequently
\begin{align}
G(A^{g_1}, g_2) &= I - i\Lambda_\beta (A+D\alpha) - \fract12
\Lambda_\beta\star
\Lambda_\beta - i\Lambda_\beta^{(2)}\label{e3.16c}\\
\begin{split}
G(A, g_{12}) &= I - i\Lambda_{\alpha+\beta-\frac i2[\alpha,\beta]}
  - \fract12 \Lambda_{\alpha+\beta}\star
\Lambda_{\alpha+\beta} - i\Lambda_{\alpha+\beta}^{(2)} \\
 &= I - i\Lambda_\alpha  - i\Lambda_\beta    -\fract 12\Lambda_{[\alpha,\beta]} 
 -\fract12 (\Lambda_\alpha + \Lambda_\beta)\star  (\Lambda_\alpha +
\Lambda_\beta)   - i\Lambda_{\alpha+\beta}^{(2)}\ .   \label{e3.16d}
\end{split}
\end{align}
\end{subequations}
The second equality in \refeq{e3.16d} follows from the first by by linearity:
$\Lambda_{\alpha+\beta} =  \Lambda_\alpha + \Lambda_\beta$ etc. After
rearrangements, it follows from \refeq{e3.11} that 
\numeq{e3.17}{
\begin{aligned}
%\begin{split}
\delta_\alpha\Lambda_\beta 
&\equiv \Lambda_\beta (A+D\alpha) -\Lambda_\beta\\
 &=
%\Lambda_{\alpha+\beta-\frac i2\alpha^2 - i\alpha\beta -
%\fract i2\beta^2} - \Lambda_\alpha 
%- \fract i2 \Lambda_{\alpha+\beta}\star\Lambda_{\alpha+\beta}
%+ \fract i2 \Lambda_\alpha\star \Lambda_\alpha\\
%&\qquad{} +
%\fract i2 \Lambda_\beta\star \Lambda_\beta
%  + i \Lambda_\alpha\star \Lambda_\beta
%  + \Lambda_{\alpha+\beta}^{(2)} - \Lambda_\alpha^{(2)}  - 
%\Lambda_\beta^{(2)}\\ 
% &= \Lambda_\beta - i\Lambda_{\alpha\beta} - \fract i2
%\Lambda_{\alpha^2+\beta^2} + \fract i2
 -\fract i2\Lambda_{[\alpha,\beta]} + \fract i2[\Lambda_\alpha,\Lambda_\beta]_\star 
- \Lambda_\alpha^{(2)} 
- \Lambda_\beta^{(2)} 
+ \Lambda_{\alpha+\beta}^{(2)} \ .
%\end{split}
\end{aligned}
}
Taking the portion of \refeq{e3.17} that is antisymmetric in
$\alpha\leftrightarrow\beta$ leaves
\numeq{e3.17b}{
\delta_\alpha \Lambda_\beta  - \delta_\beta\Lambda_\alpha = 
-i\Lambda_{[\alpha,\beta]} + i [\Lambda_\alpha,\Lambda_\beta]_\star \ .
}
This reproduces \refeq{e3.5}. Note that it was not necessary to introduce
auxiliary fields in the fundamental representation to arrive at the \cc. 
\bigskip\bigskip

\noindent
\emph{Added Note}: We have learned that similar results are contained in a preprint by
B.~Jur\v co, P.~Schupp, and J.~Wess, LMU-TPW~01-06, hep-th/0106110.

\end{document}